\documentclass[sigconf, pbalance=true]{acmart}
\AtBeginDocument{%
  }

\copyrightyear{2026}
\acmYear{2026}
\setcopyright{cc}
\setcctype{by}
\acmConference[SIGIR '26]{Proceedings of the 49th International ACM SIGIR Conference on Research and Development in Information Retrieval}{July 20--24, 2026}{Melbourne, VIC, Australia}
\acmBooktitle{Proceedings of the 49th International ACM SIGIR Conference on Research and Development in Information Retrieval (SIGIR '26), July 20--24, 2026, Melbourne, VIC, Australia}
\acmDOI{10.1145/3805712.3808570}
\acmISBN{979-8-4007-2599-9/2026/07}

\usepackage{fnpct}

\usepackage{enumitem}
\usepackage{cleveref}
\usepackage{multirow}
\usepackage{subcaption}
\usepackage[frozencache=true,cachedir=./]{minted2} 
\usepackage{ragged2e}
\definecolor{LightGray}{gray}{0.9}
\PassOptionsToPackage{dvipsnames}{xcolor}

\begin{document}

\title{Lighting the Way for BRIGHT: Reproducible Baselines with Anserini, Pyserini, and RankLLM}


\author{Sahel Sharifymoghaddam}
\authornote{Both authors contributed equally to this research.}
\orcid{0009-0008-8337-6930}
\affiliation{%
  \institution{University of Waterloo \\ David R. Cheriton School of Computer Science}
  \city{Waterloo}
  \country{Canada}
}
\email{sahel.sharifymoghaddam@uwaterloo.ca}

\author{Yijun Ge}
\authornotemark[1]
\orcid{0009-0002-2684-0692}
\affiliation{%
  \institution{University of Waterloo \\ David R. Cheriton School of Computer Science}
  \city{Waterloo}
  \country{Canada}
}
\email{l2ge@uwaterloo.ca}

\author{Raghav Vasudeva}
\orcid{0009-0003-3868-9886}
\affiliation{%
  \institution{University of Waterloo \\ David R. Cheriton School of Computer Science}
  \city{Waterloo}
  \country{Canada}
}
\email{r2vasude@uwaterloo.ca}

\author{Jimmy Lin}
\orcid{0000-0002-0661-7189}
\affiliation{%
  \institution{University of Waterloo \\ David R. Cheriton School of Computer Science}
  \city{Waterloo}
  \country{Canada}
}
\email{jimmylin@uwaterloo.ca}

\renewcommand{\shortauthors}{Sahel Sharifymoghaddam, Yijun Ge, Raghav Vasudeva, and Jimmy Lin}

\begin{abstract}
Retrieval benchmarks for large language models (LLMs) should reflect the long, reasoning-intensive queries typical of retrieval-augmented generation (RAG).
We present a systematic study of BRIGHT, a reasoning-focused retrieval benchmark, along with strong, reproducible reference methods integrated into Anserini, Pyserini, and RankLLM.
We evaluate lexical, sparse, dense, and fusion-based retrievers, as well as LLM rerankers, under long-query settings.
In reproducing BRIGHT’s lexical baseline, we identify a key under-documented detail: query-side BM25 (BM25Q), which applies BM25 weighting to the query itself.
On long, multi-sentence queries, BM25Q consistently outperforms standard BM25, making it the strongest lexical baseline for reasoning-oriented retrieval.
We further audit the BRIGHT corpus, uncovering data quality issues that impact evaluation, and offer mitigation.
Finally, we study the generalizability of BM25Q across five additional benchmarks, finding its gains largely specific to BRIGHT, while fusion with standard BM25 provides the most consistent improvements across datasets.
\end{abstract}




\maketitle

\section{Introduction}

The growing use of large language models (LLMs) and retrieval-augmented generation (RAG)~\cite{lewis2020retrieval, ram2023context, rag_survey} is transforming the role of information retrieval. In RAG, an LLM conditions on retrieved documents to generate more accurate and grounded responses, leveraging both proprietary and publicly available data. The BRIGHT benchmark~\cite{bright} was developed to evaluate retrieval systems specifically for this application, focusing on their ability to handle the complex, reasoning-intensive queries often found in user prompts. 

To facilitate further research with the BRIGHT benchmark, we have integrated reproducible baselines into two popular information retrieval (IR) toolkits, Anserini~\cite{anserini, anserini2} and Pyserini~\cite{pyserini}, for convenient first-stage retrieval. The baselines we incorporate include BM25~\cite{bm25}, a classic lexical retrieval algorithm, and BGE-large-en-v1.5~\cite{bge}, a representative dense retrieval model, both from the original BRIGHT work.
We also include SPLADE-v3~\cite{spladev3}, a representative model for learned sparse retrieval.
Finally, we include Diver-Retriever-4B~\cite{diver} and Reason-Embed-4B~\cite{reasonembed}, two of the top-performing reasoning-aware retrievers on the BRIGHT leaderboard.
Additionally, we have incorporated BRIGHT into RankLLM~\cite{rankllm}, a widely used Python package for second-stage reranking with LLMs, which has been shown to provide significant improvements in retrieval effectiveness~\cite{LRL, RankGPT, rankzephyr, first}.

In establishing these baselines, we find that the BM25 results reported in the original BRIGHT paper~\cite{bright} differ from what we obtained initially using Anserini and Pyserini. Like Lucene~\cite{lucene}, by default, Anserini uses bag-of-words (BoW) to generate sparse query vectors, whereas the implementation used by the BRIGHT paper applies the BM25 scoring algorithm to each query token to obtain token weights, a variant we refer to as ``query-side BM25'' (BM25Q).

Finally, we investigate the generality of the BM25Q technique and assess whether its benefits extend beyond BRIGHT to other standard retrieval benchmarks. To summarize, our five primary contributions are:
\begin{itemize}[leftmargin=*]

    \item \textbf{BM25 vs.\ BM25Q, Analysis and Insights (New).} We isolate and implement both standard BM25 (BoW queries) and BM25Q, enabling a controlled comparison. We show that BM25Q's improvements are driven by long query effects and characterize when it outperforms standard BM25.
    \item \textbf{Reproducible BRIGHT Baselines (Reproducibility).} We provide fully reproducible implementations of BRIGHT retrieval and reranking pipelines, integrating first-stage baselines into Anserini/Pyserini and enabling listwise reranking via RankLLM.
    \item \textbf{Fusion Strategies on BRIGHT (New).} We present a systematic evaluation of hybrid fusion methods (RRF and NAF) on BRIGHT, demonstrating that simple fusion strategies with BM25Q can surpass the effectiveness of individual retrievers, including widely used neural models.
    \item \textbf{BRIGHT Corpus Audit (New).} We conduct the first systematic audit of the BRIGHT corpora, identifying duplicates and degenerate passages, and quantify their impact on retrieval evaluation.
    \item \textbf{Generalization Beyond BRIGHT (New):} We evaluate BM25, BM25Q, and fusion strategies across five additional benchmark suites, showing that BM25Q gains are not universal and depend strongly on query characteristics.

\end{itemize}

\section{Background and Related Work}

\paragraph{BRIGHT} This is a benchmark designed to evaluate how well retrieval systems handle reasoning-intensive tasks, which remains a significant challenge for current state-of-the-art methods~\cite{bright}.
The benchmark consists of multiple heterogeneous corpora spanning several domains, including Wikipedia-style passages, technical and scientific documents, and web-derived content. 
Overall, BRIGHT contains approximately tens of thousands to hundreds of thousands of documents per corpus and a few thousand queries distributed across tasks, with each query requiring retrieval based on multi-hop reasoning, semantic composition, or document-level understanding rather than simple keyword matching or semantic similarity.
Further details on the benchmark construction, task formulation, and dataset statistics can be found in the original paper~\cite{bright}.

Unlike standard IR benchmarks such as BEIR~\cite{beir} and TREC Deep Learning tracks~\cite{dl19,dl20,dl21,dl22,dl23}, where queries are typically short keyword-like expressions, BRIGHT queries are significantly longer and often resemble natural-language questions or problem statements.
Many queries require combining multiple pieces of evidence or reasoning across concepts, making purely lexical matching or shallow semantic similarity insufficient. 
For example, a query may ask for a theorem that can be applied to solve a given mathematical problem. Answering such a query requires understanding the structure of the problem, identifying relevant mathematical principles, and mapping them to applicable theorems, rather than relying on surface-level lexical overlap or semantic similarity alone.

This difficulty is reflected in BRIGHT's leaderboard,\footnote{\url{https://brightbenchmark.github.io/}}  where the highest nDCG@10 scores remain relatively low. A common strategy to improve effectiveness is the use of LLMs, either for query expansion or reranking. This pattern is evident in the leaderboard, where top-performing systems consistently incorporate LLM-based components; notably, systems using LLM-based query expansion achieve nDCG@10 scores more than 10 percentage points higher than those using the original queries~\cite{diver}. 
Although BRIGHT offers both original and LLM-expanded queries along with first-stage retrieval runs, the results for the original queries are significantly weaker than those for the LLM-enhanced queries; nevertheless, they remain an essential part of the retrieval pipeline.
In this work, we reproduce two widely used first-stage retrieval baselines from the BRIGHT paper using BM25 and BGE-large-en-v1.5 with the original queries.

\paragraph{BM25} This is a family of probabilistic bag-of-words (BoW) ranking functions \citep{bm25} that originated in Okapi's TREC-3~\citep{trec3} runs as a compromise between BM11 and BM15 versions to perform lexical matching with sensible term frequency (TF) saturation and length normalization~\citep{robertson1995okapi}. A widely used BM25 version scores a document \(D\) for a query \(Q=\{q_i\}\) as:
\begin{equation}
\label{eq:bm25}
\mathrm{score}(D,Q)=\sum_{i}
\mathrm{IDF}(q_i)\;
\frac{(k_1+1)\,tf_{i,D}}{tf_{i,D}+k_1\,\ell(D)}\;
\frac{(k_3+1)\,qtf_i}{k_3+qtf_i},
\end{equation}
where \(tf_{i,D}\) is the frequency of \(q_i\) in \(D\); \(\ell(D)=1-b+b\,|D|/\overline{|D|}\) is the length-normalization term; and \(IDF(q_i)\) is the inverse document frequency defined as follows (with \(N\) and \(n_i\) referring to the total number of documents and those containing term \(q_i\)):
\begin{equation}
\label{eq:idf}
\mathrm{IDF}(q_i)=log
\frac{N - n_i + 0.5}{n_i + 0.5}\;
\end{equation}

In this formulation, \(k_1\) controls the influence of term frequency saturation, \(b\) determines the strength of document length normalization, and $k_3$ controls how query term frequency (\textit{qtf}) contributes to the retrieval score. When $k_3=\infty$, the model reduces to a bag-of-words representation in which term weights increase linearly with frequency, whereas $k_3=0$ yields a binary representation that considers only term presence. Early BM25 variants often omitted query term frequency, but later formulations introduced finite values of $k_3$ to allow repeated query terms to influence scoring~\cite{robertson1995okapi}. Many modern implementations today adopt the bag-of-words setting, treating \textit{qtf} linearly and thereby allowing repeated terms to affect retrieval scores~\cite{whichBM25,anserini,lucene}.

While BM25-style term weighting is well established on the document side, its application on the query side has received comparatively limited attention, with some notable exceptions.
Prior work~\cite{bm25q} generally assumes that query term frequency has limited impact, reflecting the short length and low repetition typical of keyword queries.
Alternative lines of work have explored query reweighting and expansion approaches, such as divergence-from-randomness models~\cite{dfr}, which adjust query term importance based on statistics from an initial set of retrieved documents.
However, these approaches rely on a first-pass retrieval stage, whereas BM25Q applies term-frequency saturation directly to the query within a single-stage scoring function.

For long queries---such as in document similarity tasks where the query is itself another document~\cite{lin2007pubmed}---both term-frequency effects and query length normalization become important.
This setting is particularly relevant for BRIGHT, where queries are substantially longer than in traditional datasets and frequently contain repeated tokens.
To account for these characteristics, the authors of the BRIGHT paper apply BM25-style term-frequency saturation symmetrically to both documents and queries in their baseline formulation, resulting in a ``query-as-document'' factorization:

\begin{equation}
\label{eq:bm25q}
\mathrm{score}(D,Q)
= \sum_{t\in D\cap Q}\!\mathrm{IDF}(t)^{2}\,
\frac{(k_1+1)\, tf_{t,D}}{tf_{t,D}+k_1\,\ell(D)}\;
\frac{(k_1+1)\, tf_{t,Q}}{tf_{t,Q}+k_1\,\ell(Q)},
\end{equation}
where \(\ell(x)=1-b+b\,|x|/\overline{|D|}\) uses the \emph{corpus} average length \(\overline{|D|}\) for length normalization on both sides. 
Compared to \cref{eq:bm25}, this formulation places \(\mathrm{IDF}(t)\) in both vectors, resulting in an \(\mathrm{IDF}(t)^2\) term that further emphasizes rare terms while tempering the influence of repeated query tokens. 
We refer to this variant as \emph{query-side BM25} (BM25Q).

\paragraph{Learned Dense and Sparse Retrieval} These methods have driven the
recent advances in information retrieval in two major directions beyond traditional lexical methods.
Dense retrieval models~\cite{dpr} encode queries and documents into continuous embedding vectors and compute similarity through inner products. 
Among these, the BGE~\cite{bge} family of models has become widely adopted due to its strong performance across diverse benchmarks.
In particular, BRIGHT reports results with BGE-large-en-v1.5, a dense model that generates high-quality semantic representations for both queries and documents, enabling retrieval based on meaning rather than exact term matches.

In parallel, learned sparse retrieval models extend the bag-of-words paradigm by introducing neural weighting mechanisms~\cite{conceptualFramework}.
SPLADE-v3~\cite{spladev3} exemplifies this approach by analyzing query and document tokens and assigning importance weights that capture both lexical and semantic relevance.
Importantly, SPLADE-v3 can also expand the representation by introducing semantically related tokens that are not present in the original text.
This enables the model to capture richer relationships while retaining the interpretability and efficiency of sparse representations.
Together, learned dense and sparse retrievers represent two complementary strategies: dense models capture deep semantic similarity in continuous space, while sparse models refine token-level weighting in discrete space. 
BRIGHT provides a valuable setting to compare these approaches, since its reasoning-intensive queries stress both semantic understanding and token-level precision.

\paragraph{Listwise Reranking}
Advances in instruction-following LLMs have made them effective listwise rerankers~\cite{RankGPT, rankzephyr}.
In this setting, reranking is formulated as sorting a list of candidate documents for a given query according to a relevance criterion, and is treated as a generative task in which the LLM produces an ordered list of candidate identifiers.
This approach leverages the model's reasoning ability to compare candidates with respect to the query, making it particularly well suited for reasoning-intensive ranking tasks such as BRIGHT.

\paragraph{Reasoning-Focused Retrievers and Rerankers}
Recent work on reasoning-intensive IR (e.g., BRIGHT) moves beyond lexical or surface-level semantic matching toward modeling task-relevant helpfulness, where a document may be relevant because it provides the missing concept, rule, or intermediate knowledge needed to solve the query rather than directly paraphrasing the answer. \cite{reasonir, reasonembed, diver, reasonrank}
A common recipe is to start from a strong general-purpose embedding or encoder model and post-train it with automatically constructed supervision that encodes reasoning-aware relevance---typically by synthesizing challenging queries from documents and pairing them with plausible-but-unhelpful hard negatives.~\cite{reasonir, reasonembed}
For reranking, many approaches similarly begin with a generic cross-encoder or LLM reranker and train it on reasoning-intensive, domain-diverse data (often labeled or filtered by a strong reasoning model), sometimes combining supervised fine-tuning with reinforcement learning or other post-training techniques to better capture multi-step relevance signals~\cite{reasonrank, diver, reasonembed}.

\begin{table}[t]\centering
\caption{Retrieval effectiveness (nDCG@10) of BM25 variants on BRIGHT ($k_1 = 0.9, b = 0.4$). We compare Bag-of-Words (BoW) and query-side BM25 (BM25Q) using default bucketized length normalization against their accurate (A.) counterparts. Original BRIGHT (Orig.) numbers are included as a baseline. All values are percentages.}
\vspace{-2mm}\label{tab:bm25}
\setlength{\tabcolsep}{2pt}
\resizebox{0.9\columnwidth}{!}{
\begin{tabular}{lccccc}\toprule
\multirow{3}{*}{\textbf{Dataset}} & \multicolumn{4}{c}{\textbf{Anserini}} & \multicolumn{1}{c}{\textbf{Orig.}} \\
\cmidrule(lr){2-5}\cmidrule(lr){6-6}

& \textbf{(a)} & \textbf{(b)} & \textbf{(c)} & \textbf{(d)} & \textbf{(e)} \\

& \textbf{BoW}
& \textbf{BoW A.}
& \textbf{BM25Q}
& \textbf{BM25Q A.}
& \textbf{BM25Q} \\
\cmidrule{1-6}
\multicolumn{6}{c}{\textbf{Stack Exchange}}\\
Biology & 18.2 & 17.5 & 19.7 & 18.9 & 18.9 \\
Earth Science & 27.9 & 27.0 & 27.9 & 27.2 & 27.2 \\
Economics & 16.5 & 15.9 & 15.2 & 14.8 & 14.9 \\
Psychology & 13.4 & 12.9 & 12.7 & 12.6 & 12.5 \\
Robotics & 10.9 & 10.7 & 13.9 & 13.7 & 13.6 \\
Stack Overflow & 16.3 & 16.6 & 18.6 & 18.5 & 18.4 \\
Sust. Living & 16.1 & 16.2 & 15.2 & 15.0 & 15.0 \\
\textbf{Average} & \textbf{17.0} & \textbf{16.7} & \textbf{17.6} & \textbf{17.2} & \textbf{17.2} \\
\midrule
\multicolumn{6}{c}{\textbf{Coding}}\\
LeetCode & 24.7 & 24.6 & 25.0 & 24.4 & 24.4 \\
Pony & 4.3 & 4.0 & 7.9 & 7.7 & 7.9 \\
\textbf{Average} & \textbf{14.5} & \textbf{14.3} & \textbf{16.5} & \textbf{16.1} & \textbf{16.2} \\
\midrule
\multicolumn{6}{c}{\textbf{Theorem}}\\
AoPS & 6.5 & 6.4 & 6.3 & 6.2 & 6.2 \\
TheoremQA - Q & 7.3 & 7.5 & 10.4 & 10.4 & 10.4  \\
TheoremQA - T & 2.1 & 2.1 & 4.9 & 4.9 & 4.9 \\
\textbf{Average} & \textbf{5.3} & \textbf{5.3} & \textbf{7.2} & \textbf{7.2} & \textbf{7.2} \\
\midrule
\midrule
\textbf{Overall Avg.} & \textbf{13.7} & \textbf{13.5} & \textbf{14.8} & \textbf{14.5} & \textbf{14.5} \\
\bottomrule
\end{tabular}
\vspace{-3mm}
} 
\end{table}

\begin{table*}[t]
\centering
\caption{First-stage retrieval effectiveness (nDCG@10; reported as $\times$100) on BRIGHT with BM25Q as the lexical baseline. Neural retrievers include generic SPLADE-v3 (S-v3) and BGE-large-en-v1.5 (BGE), as well as reasoning-focused retrievers Diver-Retriever-4B (Diver) and Reason-Embed-4B (R-Em.). Results are categorized into (1) individual retrievers, (2) RRF fusion with BM25Q, and (3) NAF fusion with BM25Q.}
\label{tab:bright_retrieval}
\setlength{\tabcolsep}{3pt}
\resizebox{0.8\textwidth}{!} {
\begin{tabular}{l  r  rrcl  rrcl  rrcl}
\toprule
\multirow{2}{*}{\textbf{Dataset}} & \textbf{Base} & \multicolumn{4}{c}{\textbf{(1) Individual Retrievers}} & \multicolumn{4}{c}{\textbf{(2) RRF with BM25Q}} & \multicolumn{4}{c}{\textbf{(3) NAF with BM25Q}} \\
\cmidrule(lr){3-6}\cmidrule(lr){7-10}\cmidrule(lr){11-14}
& \textbf{BM25Q} & \textbf{S-v3} & \textbf{BGE} & \textbf{Diver} & \textbf{R-Em.} & \textbf{S-v3} & \textbf{BGE} & \textbf{Diver} & \textbf{R-Em.}  & \textbf{S-v3} & \textbf{BGE} & \textbf{Diver} & \textbf{R-Em.}  \\
\midrule
\multicolumn{14}{c}{\textbf{Stack Exchange}} \\
Biology & 19.7 & 21.0 & 12.4 & 42.1 & 54.5 & 22.0 & 17.5 & 35.7 & 52.8 & 23.7 & 20.3 & 34.6 & 38.2 \\
Earth Science & 27.9 & 26.7 & 25.5 & 46.8 & 54.0 & 29.9 & 31.2 & 46.1 & 53.0 & 30.9 & 31.6 & 45.1 & 48.8 \\
Economics & 15.2 & 16.0 & 16.6 & 22.4 & 34.4 & 17.0 & 18.0 & 27.0 & 31.2 & 16.5 & 19.2 & 21.7 & 26.9 \\
Psychology & 12.7 & 15.3 & 18.1 & 34.4 & 46.1 & 15.8 & 18.5 & 32.9 & 43.8 & 16.5 & 18.5 & 28.7 & 34.8 \\
Robotics & 13.9 & 15.8 & 12.3 & 21.5 & 34.6 & 17.0 & 15.4 & 26.2 & 28.0 & 15.7 & 15.3 & 21.2 & 27.4 \\
Stack Overflow & 18.6 & 12.9 & 11.0 & 20.9 & 35.9 & 17.5 & 18.2 & 30.1 & 29.4 & 18.6 & 20.5 & 25.8 & 33.4 \\
Sust. Living & 15.2 & 15.0 & 14.4 & 25.1 & 37.0 & 15.2 & 15.8 & 26.5 & 31.0 & 16.4 & 16.5 & 23.2 & 28.6 \\
\textbf{Average} & \textbf{17.6} & \textbf{17.5} & \textbf{15.7} & \textbf{30.4} & \textbf{42.4} & \textbf{19.2} & \textbf{19.2} & \textbf{32.1} & \textbf{38.4} & \textbf{19.8} & \textbf{20.3} & \textbf{28.6} & \textbf{34.0} \\
\midrule
\multicolumn{14}{c}{\textbf{Coding}} \\
LeetCode & 25.0 & 26.0 & 26.7 & 37.8 & 37.1 & 28.2 & 29.3 & 40.6 & 38.0 & 28.8 & 30.1 & 39.3 & 41.9 \\
Pony & 7.9 & 14.4 & 3.4 & 12.9 & 12.0 & 16.3 & 15.7 & 14.8 & 25.6 & 15.5 & 10.6 & 15.1 & 14.7 \\
\textbf{Average} & \textbf{16.5} & \textbf{20.2} & \textbf{15.0} & \textbf{25.4} & \textbf{24.6} & \textbf{22.3} & \textbf{22.5} & \textbf{27.7} & \textbf{31.8} & \textbf{22.1} & \textbf{20.4} & \textbf{27.2} & \textbf{28.3} \\
\midrule
\multicolumn{14}{c}{\textbf{Theorem}} \\
AoPS & 6.3 & 6.9 & 6.4 & 10.3 & 11.3 & 8.3 & 7.0 & 12.9 & 12.5 & 8.6 & 7.6 & 12.3 & 13.1 \\
TheoremQA - Q & 10.4 & 11.1 & 14.1 & 37.7 & 40.7 & 11.1 & 12.7 & 25.5 & 39.9 & 11.9 & 14.5 & 30.1 & 32.1 \\
TheoremQA - T & 4.9 & 5.5 & 5.3 & 38.0 & 45.5 & 7.7 & 8.8 & 24.2 & 45.5 & 7.3 & 8.1 & 29.0 & 32.9 \\
\textbf{Average} & \textbf{7.2} & \textbf{7.9} & \textbf{8.6} & \textbf{28.6} & \textbf{32.5} & \textbf{9.0} & \textbf{9.5} & \textbf{20.9} & \textbf{32.6} & \textbf{9.3} & \textbf{10.1} & \textbf{23.8} & \textbf{26.1} \\
\midrule
\midrule
\textbf{Overall Avg.} & \textbf{14.8} & \textbf{15.6} & \textbf{13.8} & \textbf{29.1} & \textbf{36.9} & \textbf{17.2} & \textbf{17.3} & \textbf{28.5} & \textbf{35.9} & \textbf{17.5} & \textbf{17.7} & \textbf{27.2} & \textbf{31.1} \\
\bottomrule
\end{tabular}
}
\end{table*}
\section{Experimental Setup}
\label{sec:setup}

Our experiments employ Anserini~\cite{anserini,anserini2} and Pyserini~\cite{pyserini} for first-stage retrieval and  RankLLM~\cite{rankllm} for second-stage reranking, enabling fully reproducible experimental pipelines.
We select these toolkits---and provide direct integrations with them---because they are widely adopted in the IR community and support streamlined, two-click reproducibility.\footnote{\url{https://castorini.github.io/pyserini/2cr/bright.html}}
Documentation and usage instructions are available in the respective GitHub repositories,\footnote{\url{https://github.com/castorini/anserini/tree/master/docs/reproduce/from-document-collection}}\footnote{ \url{https://github.com/castorini/pyserini/tree/master/scripts/bright}}\footnote{\url{https://github.com/castorini/rank_llm/tree/main/src/rank_llm/demo}}
 and prebuilt indexes are released on HuggingFace.\footnote{\url{https://huggingface.co/datasets/castorini/prebuilt-indexes-bright}}
 \noindent For all experiments, we utilize Pyserini's default BM25 parameters ($k_1{=}0.9$, $b{=}0.4$). We abstain from tuning these parameters as the BRIGHT dataset does not provide a dedicated development split for parameter selection.
 We evaluate effectiveness using nDCG@10, which is the metric used for the BRIGHT leaderboard.
 The remainder of this section  describes the setup for each of our experiments.

\subsection{Isolating BM25 Implementation Choices}
To quantify the impact of lexical retrieval configurations on long, reasoning-heavy queries, we examine two orthogonal implementation dimensions:

\begin{itemize}[leftmargin=*]
\item \textbf{Query Representation.} 
We compare the standard \emph{bag-of-words} (BoW) implementation---where query term frequencies are treated linearly---against \emph{query-side BM25} (BM25Q). The latter applies BM25-style term-frequency saturation and length normalization to the query itself (treating it as a ``document'' per~\autoref{eq:bm25q}).

\item \textbf{Document-Length Normalization Precision.} 
Lucene's default \texttt{BM25Similarity} optimizes scoring speed by quantizing document-length norms into a single byte, using a 256-value lookup table for normalization factors~\cite{lucene, anserini}. While this approximation typically yields negligible differences in effectiveness~\cite{whichBM25}, we compare it against Anserini's exact-length implementation to assess its impact on reproducibility. To ensure a faithful comparison with the BRIGHT baseline, we include the exact-length version for reference but maintain the Lucene default for consistency with broader Anserini benchmarks.
\end{itemize}
Both of these configurations can be toggled via flags in the indexing and retrieval commands.

\begin{figure}[tb]
    \centering
    \begin{subfigure}[t]{\columnwidth}
        \centering
        \includegraphics[width=\columnwidth]{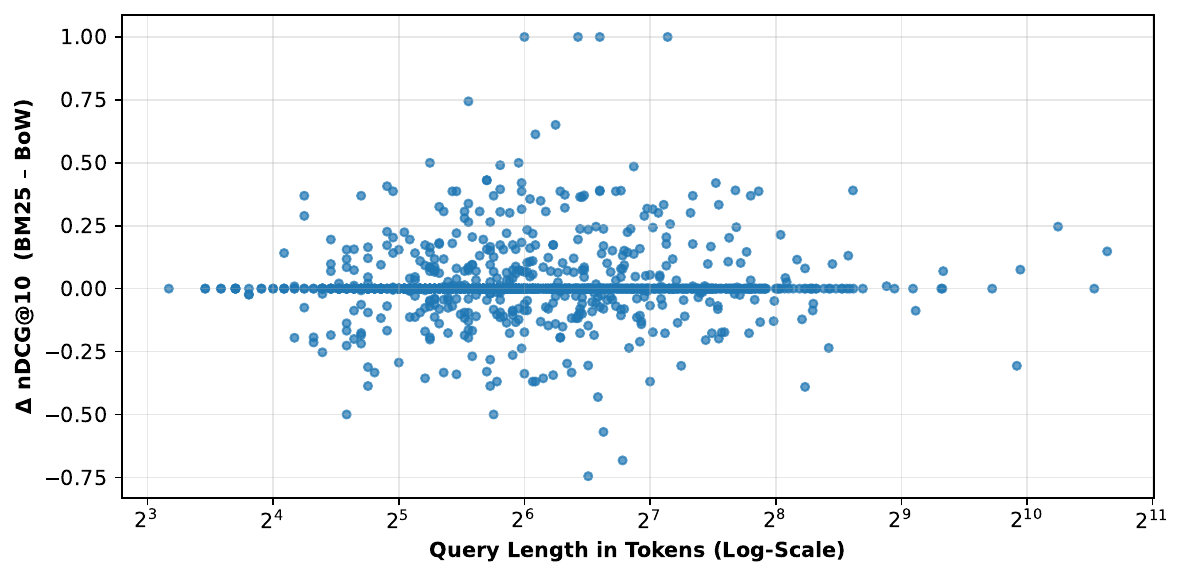}
    \end{subfigure}%
    \hfill
    \begin{subfigure}[t]{\columnwidth}
        \centering
        \includegraphics[width=\columnwidth]{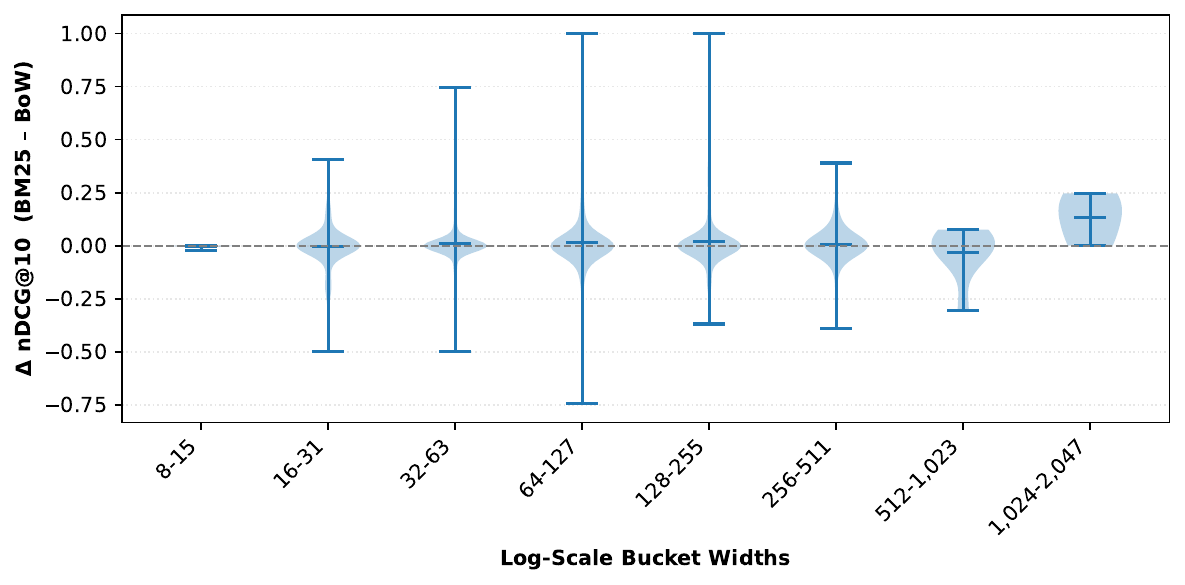}
    \end{subfigure}
\vspace{-2mm}
    \caption{Query length (in tokens) vs. $\Delta$nDCG@10 (BM25Q - BoW) for all queries in BRIGHT tasks.}
    \label{fig:aggregated_task_query_analysis}
\end{figure}

\subsection{BRIGHT Reference Methods}
We establish reproducible reference methods for BRIGHT spanning three paradigms to evaluate lexical, semantic, and reasoning-centric capabilities in different stages of the retrieval pipeline:

\begin{itemize}[leftmargin=*]
    \item \textbf{First-Stage Retrieval.} We evaluate representative retrievers from major families: BM25 (lexical), BGE-large-en-v1.5 (dense), and SPLADE-v3 (learned sparse). Furthermore, we include Diver-4B and Reason-Embed-4B, which utilize reasoning-enhanced contrastive fine-tuning and currently rank among the top-3 on the BRIGHT leaderboard as of February 12, 2026.
    
    \item \textbf{Second-Stage Reranking.} We rerank the top-$100$ candidates using listwise rerankers via the RankLLM framework. This includes out-of-the-box open-source models---Qwen3-8B and gpt-oss-120b---as well as ReasonRank-32B, a model specifically optimized for reasoning-intensive listwise reranking through reinforcement learning from ranking-based rewards. To accommodate BRIGHT's long-form queries and document snippets, we set the reranking context window to 16k tokens. We use the default window size of 20 and stride of 10 for the sliding window algorithm~\cite{RankGPT}. For ReasonRank, we use their original ranking prompt~\footnote{\url{https://github.com/8421BCD/ReasonRank}} as is~\cite{reasonrank}. The prompt used for the out-of-the-box models is shown in~\autoref{fig:reranking-prompt}. Here, inspired by the ReasonEmbed's custom query instructions per BRIGHT task, we have slightly altered the default RankZephyr~\cite{rankzephyr} prompt template to include task-specific relevance criteria.
    \item \textbf{Fusion.} To leverage complementary retrieval signals, we consider both \emph{Reciprocal Rank Fusion} (RRF) and \emph{Normalized Average Fusion} (NAF) of individual retriever results with BM25Q. For the second-stage reranking input, we adopt a ``best-of-first-stage'' strategy: we compare the effectiveness of individual retrievers against their fused combinations and select the result with the best nDCG@10 (solo or fused) to serve as the candidate set for RankLLM. This ensures that for each retriever, the reranker is evaluated on the most promising pool of candidates available from the first-stage.
\end{itemize}

\subsection{Generalizability Study Beyond BRIGHT}
To test whether BM25Q's behavior is specific to BRIGHT or persists across standard IR workloads, we conduct a head-to-head comparison of BoW, BM25Q, and their tuning-light fusions---RRF and NAF---across five additional benchmark suites: BEIR~\citep{beir}, CURE v1~\citep{cure}, TREC Disks 1 \& 2~\citep{trec1,trec2,trec3}, MIRACL~\citep{miracl}, and TREC DL19--DL23 tracks~\cite{dl19,dl20,dl21,dl22,dl23} with MS MARCO passage retrieval (v1 and v2)~\citep{msmarco}.
For this experiment, we look at the statistical significance of the results via bootstrap sampling. Exact permutations are used for per-suite tests, while Monte Carlo sampling is used for the all-tasks pooled analysis.
Using this technique, we compute 95\% \textit{confidence intervals} (CIs) and treat $p>0.005$ as \textit{not statistically significant} (NS).
\begin{table}[t]\centering
\caption{Document and query length statistics (token counts measured with the Pyserini analyzer~\cite{pyserini}) for BRIGHT.}\vspace{-1mm}\label{tab:document_query_length}
\setlength{\tabcolsep}{3pt}
\resizebox{\columnwidth}{!}{ 
\begin{tabular}{lrrrrrrrr}\toprule
\multirow{2}{*}{\textbf{Dataset}} & \multicolumn{4}{c}{\textbf{Document Length}} & \multicolumn{4}{c}{\textbf{Query Length}} \\
\cmidrule(lr){2-5}\cmidrule(lr){6-9}
&\textbf{min} &\textbf{max} &\textbf{mean} &\textbf{stdev} 
&\textbf{min} &\textbf{max} &\textbf{mean} &\textbf{stdev} \\\midrule

\multicolumn{9}{c}{\textbf{Stack Exchange}}\\
Biology &0 &3,325 &37.2 &46.4 
&14 &301 &63.1 &44.4 \\
Earth Science &0 &25,416 &40.6 &163.8 
&9 &179 &55.2 &34.2\\
Economics &0 &4,138 &44.0 &106.7 
&21 &241 &87.9 &50.4\\
Psychology &0 &26,064 &41.8 &190.6 
&15 &237 &78.5 &44.1 \\
Robotics &0 &3,692 &31.7 &73.0 
&19 &1,594 &199.8 &282.3 \\
Stack Overflow &0 &843 &117.8 &143.2 
&21 &845 &135.7 &103.5 \\
Sust. Living &0 &18,756 &38.8 &143.2 
&15 &309 &81.1 &54.3 \\

\midrule
\multicolumn{9}{c}{\textbf{Coding}}\\
LeetCode &7 &12,231 &85.4 &110.0 
&53 &493 &176.0 &77.2 \\
Pony &0 &246 &33.5 &24.0 
&21 &117 &45.6 &18.2 \\

\midrule
\multicolumn{9}{c}{\textbf{Theorem}}\\
AoPS &8 &1,125 &110.1 &82.5 
&12 &195 &43.6 &23.9 \\
TheoremQA - Q &8 &1,125 &110.1 &82.5 
&11 &141 &51.5 &19.3\\
TheoremQA - T &7 &2,706 &118.4 &128.2 
&23 &91 &50.9 &16.7\\
\bottomrule
\end{tabular}
}
\end{table}

\section{Experimental Results}

\subsection{BM25 Implementation Analysis}
\autoref{tab:bm25} shows nDCG@10 for different variations of BM25 implementations. Notably, we are able to replicate the BM25 results from the original BRIGHT paper almost exactly using the Anserini toolkit; see columns (d) and (e).

\paragraph{Quantized vs. Accurate Length Normalization.} 
Looking at columns (a) vs.~(b) and~(c) vs.~(d) in \autoref{tab:bm25}, we find that quantized and exact document-length normalization yield nearly identical results. On average, the quantized variant, which is the Lucene default, scores slightly higher by approximately 0.2–0.3 nDCG@10 \textit{percentage points} (PP). Analysis of the document statistics in \autoref{tab:document_query_length} reveals two general patterns:
\begin{itemize}[leftmargin=*]
    \item \textbf{Discrimination Loss:} For tasks where document lengths are tightly clustered, such as Stack Overflow and TheoremQA - Q, accurate normalization performs slightly better. This is because Lucene's single-byte quantization collapses nearby lengths into the same bucket, reducing the model's ability to discriminate between documents.
    \item \textbf{Penalty Mitigation:} In collections with a heavy long tail of documents, such as Biology or Earth Science, quantization often helps by softening the BM25 length penalty. This mitigates the tendency to over-penalize extremely long documents, acting as a form of implicit regularization.
\end{itemize}
These findings support existing conclusions that quantization has a negligible impact on overall effectiveness. We utilize the quantized variant moving forward for consistency with Anserini defaults.

\begin{table*}[t]
\centering
\caption{Second-stage reranking effectiveness (nDCG@10, $\times$100) on BRIGHT with SPLADE-v3 (S-v3) and BGE-large-en-v1.5 (BGE) as generic retrievers with BM25Q NAF fusions, and reasoning-focused Diver-Retriever-4B (Diver) and Reason-Embed-4B (R-Em). Top-100 candidates are reranked using Qwen3-8B (Qw.), gpt-oss-120b (GPT), and ReasonRank-32B (R-Ra) rerankers.}
\label{tab:reranking_results}
\setlength{\tabcolsep}{3pt}
\resizebox{0.9\textwidth}{!}{
\begin{tabular}{l rrcl rrcl rrcl rrcl}
\toprule
\multirow{2}{*}{\textbf{Dataset}} & \multicolumn{4}{c}{\textbf{S-v3 (w. BM25Q NAF)}} & \multicolumn{4}{c}{\textbf{BGE (w. BM25Q NAF)}} & \multicolumn{4}{c}{\textbf{Diver}} & \multicolumn{4}{c}{\textbf{R-Em.}} \\
\cmidrule(lr){2-5} \cmidrule(lr){6-9} \cmidrule(lr){10-13} \cmidrule(lr){14-17}
 & \textbf{Base} & \textbf{Qw.} & \textbf{GPT} & \textbf{R-Ra.} & \textbf{Base} & \textbf{Qw.} & \textbf{GPT} & \textbf{R-Ra.} & \textbf{Base} & \textbf{Qw.} & \textbf{GPT} & \textbf{R-Ra.} & \textbf{Base} & \textbf{Qw.} & \textbf{GPT} & \textbf{R-Ra.} \\ \midrule
\multicolumn{17}{c}{\textbf{Stack Exchange}} \\
Biology & 23.7 & 41.6 & 44.0 & 42.4 & 20.3 & 36.6 & 43.1 & 40.1 & 42.1 & 53.8 & 54.9 & 56.1 & 54.5 & 55.6 & 55.7 & 58.4 \\
Earth Science & 30.9 & 45.4 & 48.9 & 47.4 & 31.6 & 48.2 & 46.5 & 48.2 & 46.8 & 50.5 & 53.5 & 51.6 & 54.0 & 50.0 & 54.5 & 51.3 \\
Economics & 16.5 & 27.8 & 32.9 & 31.2 & 19.2 & 28.0 & 33.0 & 34.5 & 22.4 & 31.4 & 38.5 & 35.3 & 34.4 & 34.5 & 39.1 & 40.0 \\
Psychology & 16.5 & 38.7 & 40.6 & 41.3 & 18.5 & 39.2 & 36.6 & 39.2 & 34.4 & 48.6 & 51.6 & 53.9 & 46.1 & 47.9 & 52.6 & 51.8 \\
Robotics & 15.7 & 32.3 & 36.9 & 36.4 & 15.3 & 25.2 & 33.7 & 30.5 & 21.5 & 27.1 & 29.9 & 34.9 & 34.6 & 32.8 & 37.0 & 39.2 \\
Stack Overflow & 18.6 & 29.0 & 31.7 & 30.7 & 20.5 & 30.2 & 32.0 & 34.1 & 20.9 & 28.1 & 34.2 & 33.4 & 35.9 & 32.7 & 38.6 & 37.4 \\
Sust. Living & 16.4 & 34.7 & 40.7 & 37.8 & 16.5 & 33.0 & 37.9 & 39.1 & 25.1 & 40.0 & 45.2 & 42.7 & 37.0 & 44.3 & 49.0 & 45.6 \\
\textbf{Average} & \textbf{19.8} & \textbf{35.6} & \textbf{39.4} & \textbf{38.2} & \textbf{20.3} & \textbf{34.3} & \textbf{37.5} & \textbf{38.0} & \textbf{30.4} & \textbf{39.9} & \textbf{44.0} & \textbf{44.0} & \textbf{42.4} & \textbf{42.5} & \textbf{46.6} & \textbf{46.2} \\ \midrule
\multicolumn{17}{c}{\textbf{Coding}} \\
LeetCode & 28.8 & 27.1 & 32.7 & 25.3 & 30.1 & 28.1 & 33.8 & 28.3 & 37.8 & 20.6 & 26.0 & 19.8 & 37.1 & 19.4 & 25.3 & 21.1 \\
Pony & 15.5 & 20.1 & 35.7 & 23.5 & 10.6 & 14.6 & 29.3 & 16.2 & 12.9 & 22.8 & 32.6 & 20.8 & 12.0 & 12.6 & 28.5 & 13.9 \\
\textbf{Average} & \textbf{22.1} & \textbf{23.6} & \textbf{34.2} & \textbf{24.4} & \textbf{20.4} & \textbf{21.4} & \textbf{31.6} & \textbf{22.3} & \textbf{25.4} & \textbf{21.7} & \textbf{29.3} & \textbf{20.3} & \textbf{24.6} & \textbf{16.0} & \textbf{26.9} & \textbf{17.5} \\ \midrule
\multicolumn{17}{c}{\textbf{Theorem}} \\
AoPS & 8.6 & 11.2 & 11.9 & 12.0 & 7.6 & 8.7 & 11.6 & 8.7 & 10.3 & 10.2 & 11.6 & 10.2 & 11.3 & 9.4 & 12.5 & 9.4 \\
TheoremQA - Q & 11.9 & 22.9 & 24.1 & 24.3 & 14.5 & 23.4 & 25.7 & 24.5 & 37.7 & 39.8 & 41.4 & 41.0 & 40.7 & 40.3 & 42.0 & 42.2 \\
TheoremQA - T & 7.3 & 20.6 & 20.4 & 18.3 & 8.1 & 19.8 & 19.9 & 20.6 & 38.0 & 44.6 & 49.4 & 46.3 & 45.5 & 49.3 & 48.4 & 45.8 \\
\textbf{Average} & \textbf{9.3} & \textbf{18.2} & \textbf{18.8} & \textbf{18.2} & \textbf{10.1} & \textbf{17.3} & \textbf{19.1} & \textbf{17.9} & \textbf{28.6} & \textbf{31.5} & \textbf{34.1} & \textbf{32.5} & \textbf{32.5} & \textbf{33.0} & \textbf{34.3} & \textbf{32.5} \\ \midrule \midrule
\textbf{Overall Avg.} & \textbf{17.5} & \textbf{29.3} & \textbf{33.4} & \textbf{30.9} & \textbf{17.7} & \textbf{27.9} & \textbf{31.9} & \textbf{30.3} & \textbf{29.1} & \textbf{34.8} & \textbf{39.1} & \textbf{37.2} & \textbf{36.9} & \textbf{35.7} & \textbf{40.3} & \textbf{38.0} \\
\bottomrule
\end{tabular}
}
\end{table*}

\begin{figure}[!tbh]
\tiny
\justifying
\begin{minted}[fontsize=\tiny, frame=lines, frame=single,linenos=false,breaklines,breaksymbol=,escapeinside=||,bgcolor=LightGray]{text}
system_message: "You are RankLLM, an intelligent assistant that ranks passages based on a defined criterion."

prefix:
  """I will provide you with {num} passages, each identified by an alphabetical label []. Given a Sustainable Living post as a query, rank the passages by how helpful they are for answering the query: {query}."""

body:
  """- Passage [{rank}]: {candidate}"""

suffix:
  """Search Query: {query}
  Rank the {num} passages above by their usefulness in answering the query. Include all passages and list them using their identifiers in descending order of usefulness.
  Output format: [] > [] (e.g., [B] > [A]).
  Respond only with the ranking and do not include any explanation or extra words."""

output_validation_regex: r"^\[[A-Z]\]( > \[[A-Z]\])*$"

output_extraction_regex: r"\[([A-Z])\]"
\end{minted}
\vspace*{-0.5cm}
\caption{Inference prompt for listwise reranking with out-of-the-box LLMs.}
\label{fig:reranking-prompt}
\end{figure}
\paragraph{Bag-of-Words vs. BM25Q Query Representation.}
When comparing columns~(a) and~(c) in \autoref{tab:bm25}, we find that BM25Q query vectors are more effective than plain BoW vectors on \emph{seven} of the twelve datasets, BoW is better on \emph{four}, and the two approaches tie on one.
From the aggregate query-length statistics (\autoref{tab:document_query_length}) we do not observe a direct correlation between a task's query-length distribution and whether BM25Q or BoW is superior.
However, a per-query analysis (\autoref{fig:aggregated_task_query_analysis}) reveals three consistent patterns:
\begin{itemize}[leftmargin=*]
\item For \emph{short} queries (\(<\!16\) tokens) the two methods are almost identical, with BoW only rarely ahead. This is consistent with the traditional practice of \emph{not} applying BM25 to queries~\cite{robertson1995okapi, robertson1996okapi}.
\item As the query length grows, both the \emph{frequency} and the \emph{magnitude} of the differences increase up to a peak, then taper off; the decline occurs because the length-normalization term in BM25 begins to dominate for very long queries. %
\item Either method can win depending on whether down-weighting frequent terms in favor of rarer ones helps or harms the query's key terms. %
As BM25Q's wins are both \emph{larger} and \emph{more numerous} at medium lengths (~\autoref{fig:aggregated_task_query_analysis}), we recommend applying BM25Q weighting to queries of roughly 16--256 tokens.
\end{itemize}

\subsection{BRIGHT Reference Results}
\paragraph{Retrieval.}
\autoref{tab:bright_retrieval} summarizes first-stage retrieval effectiveness (nDCG@10; reported as \(\times 100\)) on BRIGHT using query-side BM25 as the lexical baseline, with SPLADE-v3 (S-v3) and BGE-large-en-v1.5 (BGE) as generic neural retrievers and Diver-Retriever-4B (Diver) and Reason-Embed-4B (R-Em.) as reasoning-focused retrievers. Results are organized into (1) individual retrievers, (2) RRF fusion with BM25Q, and (3) NAF fusion with BM25Q.
To ensure the technical integrity of our experimental setup, we compare the BM25 and BGE results from the original BRIGHT paper, as well as the Diver and R-Em. results from their respective original works. While all other results in \autoref{tab:bright_retrieval} are new, these four reproduced baselines demonstrate strong agreement with published figures, showing a maximum absolute difference of only 0.3 PP. This close alignment validates the correctness of our implementation and provides a reliable foundation for our subsequent fusion analysis.

In the \emph{individual} setting (group~1), generic retrievers exhibit category-dependent strengths rather than a uniform winner. On Stack Exchange, BGE underperforms both BM25 and S-v3 on average (15.7 vs.\ 17.6/17.5), while S-v3 is competitive with BM25 overall (15.6 vs.\ 14.8 overall average) and clearly strongest among the generic models in Coding (20.2 vs.\ 16.5 BM25 and 15.0 BGE).
As expected, the reasoning retrievers are substantially stronger than all generic baselines across categories: Diver and R-Em. achieve overall averages of 29.1 and 36.9, respectively, with R-Em. leading Diver in every subset except Coding, where they are nearly tied on average (25.4 vs.\ 24.6).

Fusion with BM25Q reveals a clear asymmetry between generic and reasoning retrievers. For S-v3 and BGE, \emph{both} fusion schemes consistently improve over the corresponding individual retriever,
indicating complementary lexical--neural signals when the components are of comparable strength.
In contrast, fusing BM25Q with reasoning retrievers generally \emph{reduces} effectiveness relative to Diver and R-Em. alone, with the notable exception of Coding (where lexical cues such as API functions/library names and exact identifiers can rescue misses).

\begin{table}[t]\centering
\caption{Per-task counts of unique, short (<5 tokens), and zero-token document chunks in the BRIGHT dataset.}\vspace{-2mm}\label{tab:duplicate_docs}
\setlength{\tabcolsep}{3pt}
\resizebox{\columnwidth}{!}{ 
\begin{tabular}{lrrrrrrr}\toprule
\multirow{2}{*}{\textbf{Dataset}} & \multicolumn{1}{c}{} & \multicolumn{2}{c}{\textbf{Unique}} & \multicolumn{2}{c}{\textbf{Short}} & \multicolumn{2}{c}{\textbf{Zero Length}}\\
\cmidrule(lr){3-4}\cmidrule(lr){5-6}\cmidrule(lr){7-8}
 &\textbf{Total} &\textbf{Count} &  \textbf{(\%)} &\textbf{Count} & \textbf{(\%)} &\textbf{Count} & \textbf{(\%)} \\\midrule
\multicolumn{8}{c}{\textbf{Stack Exchange}}\\
Biology &57,359 &49,434 &86.2\% &7,534 &13.1\% &329 &0.6\% \\
Earth Science &121,249 &117,633 &97.0\% &5,182 &4.3\% &44 &0.0\% \\
Economics &50,220 &40,594 &80.8\% &13,357 &26.6\% &748 &1.5\% \\
Psychology &52,835 &43,756 &82.8\% &13,802 &26.1\% &780 &1.5\% \\
Robotics &61,961 &40,431 &65.3\% &22,617 &36.5\% &2,229 &3.6\% \\
Stack Overflow &107,081 &66,270 &61.9\% &15,749 &14.7\% &1,006 &0.9\% \\
Sust. Living &60,792 &50,142 &82.5\% &15,777 &26.0\% &613 &1.0\% \\
\midrule
\multicolumn{8}{c}{\textbf{Coding}}\\
LeetCode &413,932 &413,932 &100.0\% &0 &0.0\% &0 &0.0\% \\
Pony &7,894 &6,183 &78.3\% &98 &1.2\% &2 &0.0\% \\
\midrule
\multicolumn{8}{c}{\textbf{Theorem}}\\
AoPS &188,002 &176,508 &93.9\% &0 &0.0\% &0 &0.0\% \\
TheoremQA - Q &188,002 &176,508 &93.9\% &0 &0.0\% &0 &0.0\% \\
TheoremQA - T &23,839 &23,839 &100.0\% &0 &0.0\% &0 &0.0\% \\
\bottomrule
\end{tabular}
} 
\end{table}
Comparing fusion operators (categories 2 vs. 3), NAF tends to be slightly better for the generic retrievers, whereas RRF is markedly better for the reasoning retrievers.
A plausible explanation is that when combining similarly capable systems (e.g., BM25 and generic neural retrievers), score normalization and averaging can preserve complementary high-confidence signals from each method in the final top-10 results.
However, when one component is substantially weaker or differently calibrated (e.g., BM25 vs.\ reasoning-focused retrievers), score-based fusion can overpromote lexically plausible but semantically irrelevant candidates.
In contrast, rank-based RRF mitigates this effect by limiting the influence of any single list: documents absent from the second list are treated as ranked last $+\,1$, reducing the likelihood that spurious BM25 ``confidence'' displaces genuinely strong reasoning-based hits at the top of the ranking.

\paragraph{Reranking}
\autoref{tab:reranking_results} reports second-stage reranking effectiveness (nDCG@10$\times$100) when reranking the top-100 candidates using three listwise rerankers: Qwen3-8B (Qw.), gpt-oss-120b (GPT), and ReasonRank-32B (R-Ra.), applied on the same retrievers used in the first stage. For S-v3 and BGE, reranking is applied on their BM25Q NAF fusion results.

Reranking substantially improves effectiveness, particularly for weaker first-stage retrievers. GPT improves S-v3 from 17.5 to 33.4 (+15.9 PP) and BGE from 17.7 to 31.9 (+14.2 PP). In contrast, gains are smaller for stronger reasoning-focused retrievers: Diver improves from 29.1 to 39.1 (+10.0 PP), and Reason-Embed-4B (R-Em.) improves from 36.9 to 40.3 (+3.4 PP). This demonstrates diminishing returns from reranking as first-stage retrieval quality improves.

\begin{table}[ht]
\centering
\caption{nDCG@10 under original and adjusted qrels (with added missing gold documents) for Stack Exchange tasks. First-stage retrieval uses SPLADE-v3, BGE-large-en-v1.5, Diver-Retriever-4B, and Reason-Embed-4B as retrievers; second-stage reranking uses gpt-oss-120b.}
\label{tab:adjusted_qrels}
\setlength{\tabcolsep}{3pt}
\resizebox{\columnwidth}{!}{
\begin{tabular}{lcccccccc}
\toprule
\multirow{2}{*}{\textbf{Task}} & \multicolumn{2}{c}{\textbf{S-v3}} & \multicolumn{2}{c}{\textbf{BGE}} & \multicolumn{2}{c}{\textbf{Diver}} & \multicolumn{2}{c}{\textbf{R-Em.}} \\
\cmidrule(lr){2-3} \cmidrule(lr){4-5} \cmidrule(lr){6-7} \cmidrule(lr){8-9}
& \textbf{Orig.} & \textbf{Adj.} & \textbf{Orig.} & \textbf{Adj.} & \textbf{Orig.} & \textbf{Adj.} & \textbf{Orig.} & \textbf{Adj.} \\
\midrule
\multicolumn{9}{c}{\textbf{Retrieval}} \\

Biology        & 23.7 & 23.8 & 20.3 & 20.4 & 42.2 & 42.9 & 54.5 & 55.6 \\
Earth Science  & 30.9 & 31.1 & 31.6 & 31.8 & 46.2 & 46.5 & 54.1 & 54.5 \\
Economics      & 16.5 & 16.5 & 19.2 & 19.2 & 21.9 & 21.9 & 34.4 & 34.4 \\
Psychology     & 16.5 & 16.5 & 18.5 & 18.5 & 34.2 & 34.2 & 46.1 & 46.1 \\
Robotics       & 15.7 & 15.7 & 15.3 & 15.3 & 21.3 & 21.3 & 34.6 & 34.6 \\
Stack Overflow & 18.6 & 19.1 & 20.5 & 21.0 & 20.7 & 21.4 & 36.2 & 37.6 \\
Sust. Living   & 16.4 & 16.4 & 16.5 & 16.4 & 24.8 & 24.8 & 36.9 & 36.8 \\
\textbf{Average} & \textbf{19.8} & \textbf{19.9} & \textbf{20.3} & \textbf{20.4} & \textbf{30.2} & \textbf{30.4} & \textbf{42.4} & \textbf{42.8} \\
\midrule
\multicolumn{9}{c}{\textbf{Reranking with gpt-oss-120b}} \\

Biology        & 44.0 & 44.8 & 43.1 & 43.8 & 54.9 & 56.2 & 55.7 & 56.7 \\
Earth Science  & 48.9 & 49.1 & 46.5 & 46.7 & 53.5 & 54.1 & 54.5 & 55.1 \\
Economics      & 32.9 & 32.9 & 33.0 & 33.0 & 38.5 & 38.5 & 39.1 & 39.1 \\
Psychology     & 40.6 & 40.6 & 36.6 & 36.6 & 51.6 & 51.6 & 52.6 & 52.6 \\
Robotics       & 36.9 & 36.9 & 33.7 & 33.7 & 29.9 & 29.9 & 37.0 & 37.1 \\
Stack Overflow & 31.7 & 32.5 & 32.0 & 33.4 & 34.2 & 35.0 & 38.6 & 39.9 \\
Sust. Living   & 40.7 & 40.6 & 37.9 & 37.8 & 45.2 & 45.1 & 49.0 & 49.0 \\
\textbf{Average} & \textbf{39.4} & \textbf{39.6} & \textbf{37.5} & \textbf{37.8} & \textbf{44.0} & \textbf{44.3} & \textbf{46.6} & \textbf{47.1} \\
\bottomrule
\end{tabular}
}
\end{table}
Comparing rerankers, GPT consistently achieves the strongest overall effectiveness, yielding the highest average nDCG@10 for all retrievers. In comparison with GPT, R-Ra. is particularly effective on Stack Exchange tasks, outperforming GPT in domains such as Biology  and Robotics.
Given their comparable sizes (32B dense vs. 5.1B active parameters out of 120B MoE) this improvement suggests that R-Ra. is benefiting from its reasoning-aware training. Qw. is consistently the weakest reranker, though it still improves weaker retrievers (e.g., +11.8 PP on S-v3).

Importantly, reranking can degrade performance when applied to strong reasoning-focused retrievers using weaker rerankers. Applying Qw. to R-Em.\ reduces the average effectiveness by $-$1.2 PP, indicating that weaker rerankers may disrupt high-quality initial rankings. Sharper degradations are observed in specific domains, such as LeetCode ($-$17.7 PP).
At the domain level, the largest gains occur on reasoning-intensive tasks. For example, GPT improves Diver on TheoremQA-T by +11.4 PP and on Stack Exchange Psychology by +17.2 PP. In contrast, improvements are smaller on coding tasks, where strong retrievers already achieve competitive effectiveness. Especially for LeetCode, regardless of the LLM choice, reranking degrades the results of strong reasoning-focused retrievers.
Overall, these results show that listwise LLM reranking provides large gains for weaker retrievers but smaller gains for strong reasoning-focused retrievers, and can degrade performance when reranker capability is insufficient relative to first-stage retrieval quality. This is more common when the retriever is fine-tuned on domain-specific data such as math and coding.
\begin{table*}[t]
\caption{Average nDCG@10 and Recall@100 for BoW, BM25Q, and their fusions (RRF and NAF) across six benchmarks, together with statistical comparisons against BoW. All effectiveness values and mean differences are reported on a percentage scale (without the \% symbol). We report mean differences ($\overline{\Delta}$), 95\% confidence intervals (CI), and $p$-values for both metrics. The \emph{All tasks} row corresponds to a pooled analysis over all 75 tasks, rather than an average of the six benchmark-level averages.}
\vspace{-1mm}\label{tab:generalizability}
\setlength{\tabcolsep}{3pt}
\resizebox{0.9\textwidth}{!}{
\begin{tabular}{l l l l rr r r r rr r r}
\toprule
\multirow{2}{*}{\textbf{Dataset}} &
\multirow{2}{*}{\textbf{N}} &
\multirow{2}{*}{\textbf{Baseline}} &
\multirow{2}{*}{\textbf{New}} &
\multicolumn{2}{c}{\textbf{nDCG@10}} &
\multirow{2}{*}{\textbf{$\overline{\Delta(N-B)}$ [95\% CI]}} &
\multirow{2}{*}{\textbf{$p$}} &
\multicolumn{2}{c}{\textbf{Recall@100}} &
\multirow{2}{*}{\textbf{$\overline{\Delta(N-B)}$ [95\% CI]}} &
\multirow{2}{*}{\textbf{$p$}} \\
\cmidrule(lr){5-6}\cmidrule(lr){9-10}
& & & &
\textbf{Base} & \textbf{New} &
& &
\textbf{Base} & \textbf{New} &
& \\
\midrule

\multirow{3}{*}{BEIR} & \multirow{3}{*}{18} & BoW & BM25Q & 42.5 & 40.9 & $-$1.6 [$-$2.5, $-$0.7] & 0.002 & 58.7 & 58.1 & $-$0.6 [$-$1.2, 0.1] & 0.082 \\
& & BoW & RRF   & 42.5 & 42.1 & $-$0.4 [$-$0.8, 0.1] & 0.134 & 58.7 & 59.1 & 0.4 [0.1, 0.8] & 0.040 \\
& & BoW & NAF   & 42.5 & 42.5 & 0.0 [$-$0.5, 0.4] & 0.863 & 58.7 & 59.1 & 0.4 [0.0, 0.9] & 0.067 \\
\midrule

\multirow{3}{*}{BRIGHT} & \multirow{3}{*}{12} & BoW & BM25Q & 13.7 & 14.8 & 1.1 [0.1, 2.1] & 0.056 & 34.7 & 39.0 & 4.2 [2.2, 6.3] & 0.003 \\
& & BoW & RRF   & 13.7 & 14.7 & 1.0 [0.6, 1.4] & 0.002 & 34.7 & 38.4 & 3.7 [2.6, 4.7] & 0.001 \\
& & BoW & NAF   & 13.7 & 14.8 & 1.1 [0.7, 1.5] & 0.002 & 34.7 & 38.1 & 3.4 [2.3, 4.4] & 0.001 \\
\midrule

\multirow{3}{*}{CURE v1} & \multirow{3}{*}{10} & BoW & BM25Q & 34.3 & 33.4 & $-$0.9 [$-$1.4, $-$0.4] & 0.017 & 47.7 & 48.7 & 1.0 [0.4, 1.7] & 0.022 \\
& & BoW & RRF   & 34.3 & 34.4 & 0.1 [$-$0.2, 0.3] & 0.520 & 47.7 & 48.7 & 1.0 [0.7, 1.4] & 0.003 \\
& & BoW & NAF   & 34.3 & 34.6 & 0.3 [0.1, 0.5] & 0.050 & 47.7 & 48.9 & 1.2 [0.9, 1.6] & 0.003 \\
\midrule

\multirow{3}{*}{Disks 1 and 2} & \multirow{3}{*}{9} & BoW & BM25Q & 48.1 & 48.8 & 0.7 [$-$1.4, 2.7] & 0.536 & 18.1 & 18.9 & 0.8 [$-$0.6, 2.1] & 0.314 \\
& & BoW & RRF   & 48.1 & 49.9 & 1.8 [0.8, 2.9] & 0.006 & 18.1 & 19.0 & 0.8 [0.1, 1.6] & 0.080 \\
& & BoW & NAF   & 48.1 & 49.7 & 1.6 [0.7, 2.5] & 0.014 & 18.1 & 19.1 & 0.9 [0.1, 1.7] & 0.072 \\
\midrule

\multirow{3}{*}{MIRACL} & \multirow{3}{*}{18} & BoW & BM25Q & 38.5 & 36.3 & $-$2.2 [$-$3.4, $-$0.6] & 0.010 & 77.2 & 75.3 & $-$2.0 [$-$3.2, $-$0.6] & 0.012 \\
& & BoW & RRF   & 38.5 & 37.8 & $-$0.7 [$-$1.4, 0.2] & 0.100 & 77.2 & 77.6 & 0.4 [$-$0.2, 1.0] & 0.313 \\
& & BoW & NAF   & 38.5 & 38.2 & $-$0.3 [$-$1.0, 0.7] & 0.594 & 77.2 & 77.5 & 0.3 [$-$0.3, 1.0] & 0.489 \\
\midrule

\multirow{3}{*}{MS MARCO} & \multirow{3}{*}{8} & BoW & BM25Q & 29.6 & 28.7 & $-$0.9 [$-$1.5, $-$0.3] & 0.043 & 36.9 & 37.0 & 0.1 [$-$0.9, 1.7] & 0.961 \\
& & BoW & RRF   & 29.6 & 29.7 & 0.1 [$-$0.2, 0.4] & 0.681 & 36.9 & 37.7 & 0.8 [0.3, 1.4] & 0.019 \\
& & BoW & NAF   & 29.6 & 29.7 & 0.1 [$-$0.3, 0.7] & 0.650 & 36.9 & 37.8 & 0.9 [0.2, 1.7] & 0.027 \\
\midrule

\multirow{3}{*}{\textbf{All tasks}} & \multirow{3}{*}{\textbf{75}} & \textbf{BoW} & \textbf{BM25Q} & \textbf{35.1} & \textbf{34.3} & \textbf{$-$0.8 [$-$1.4, $-$0.2]} & \textbf{0.005} & \textbf{50.6} & \textbf{51.0} & \textbf{0.4 [$-$0.4, 1.0]} & \textbf{0.408} \\
& & \textbf{BoW} & \textbf{RRF}   & \textbf{35.1} & \textbf{35.3} & \textbf{0.2 [$-$0.2, 0.5]} & \textbf{0.441} & \textbf{50.6} & \textbf{51.7} & \textbf{1.1 [0.7, 1.5]} & \textbf{$<$ 0.001} \\
& & \textbf{BoW} & \textbf{NAF}   & \textbf{35.1} & \textbf{35.5} & \textbf{0.4 [0.0, 0.7]} & \textbf{0.026} & \textbf{50.6} & \textbf{51.7} & \textbf{1.1 [0.7, 1.5]} & \textbf{$<$ 0.001} \\
\bottomrule
\end{tabular}
}
\end{table*}

\subsection{BRIGHT Corpora}

During our experiments, we identified three data quality issues in the BRIGHT corpora.
\begin{itemize}[leftmargin=*]

\item \textbf{Duplicate Documents.}
As shown in \autoref{tab:duplicate_docs}, a substantial proportion of documents are exact duplicates after trimming leading and trailing whitespace, accounting for up to 38\% in some tasks.

\item \textbf{Very Short Documents.}
Eight of the twelve tasks contain documents with fewer than five tokens (ranging from 1.2\% to 36.5\%), and some documents are even empty. These cases typically arise when documents consist primarily of whitespace that is removed
during tokenization.\vspace{0.2em}

\leavevmode\hspace{1.2em}We therefore recommend additional cleaning of scraped documents prior to chunking them into shorter passages. This step is particularly important for reasoning-intensive retrievers fine-tuned on math and coding tasks, which are sensitive to formatting and spacing artifacts. For example, removing trailing whitespace from document \texttt{aqua\_rat\_75207} in the AoPS corpus in Pyserini reduces the cosine similarity between embeddings produced by Pyserini and the original Diver codebase\footnote{\url{https://github.com/AQ-MedAI/Diver}} from 0.9998 to 0.8700, illustrating the substantial impact that minor formatting differences can have on embedding representations.

\item \textbf{Missing Query Relevance.} Most duplicate documents correspond to low-quality fragments that are unrelated to any query and thus have limited impact on retrieval evaluation. However, in some cases, relevant gold document IDs are absent from the positive candidate lists; restoring these missing duplicates can change the measured results. To quantify the effect of missing duplicate IDs, we modified the relevance labels (qrels) as follows: for each set of identical documents, if any document ID was marked relevant for a given query, we marked all document IDs in that set as relevant.
\end{itemize}
\autoref{tab:adjusted_qrels} compares nDCG@10 scores for retrieval and reranking using the original BRIGHT qrels versus the adjusted qrels for the Stack Exchange category. The Coding and Theorem categories are excluded from this analysis, as they contain no missing gold IDs.

For first-stage retrieval, the average impact is modest, increasing nDCG@10 by 0.1--0.4 PP for different retrievers.
As expected, stronger retrievers are more heavily impacted by the original qrels; because they are more effective at surfacing relevant documents, they are disproportionately penalized when those documents are incorrectly treated as negatives. This trend is mirrored in the maximum per-task differences: 0.5 PP for BGE and S-v3, compared to 0.7 for Diver and 1.4 for R-Em.

In second-stage reranking, the average nDCG@10 discrepancies  increase slightly (0.2--0.5 PP) as the reranker further promotes truly relevant documents to the top of the list.
To mitigate this evaluation bias, we recommend removing duplicate documents from the dataset. In the meantime, we have integrated these adjusted qrels into Pyserini to support more accurate benchmarking.


\begin{figure}[tbh]
    \begin{subfigure}[t]{\columnwidth}
        \centering
        \includegraphics[width=\columnwidth]{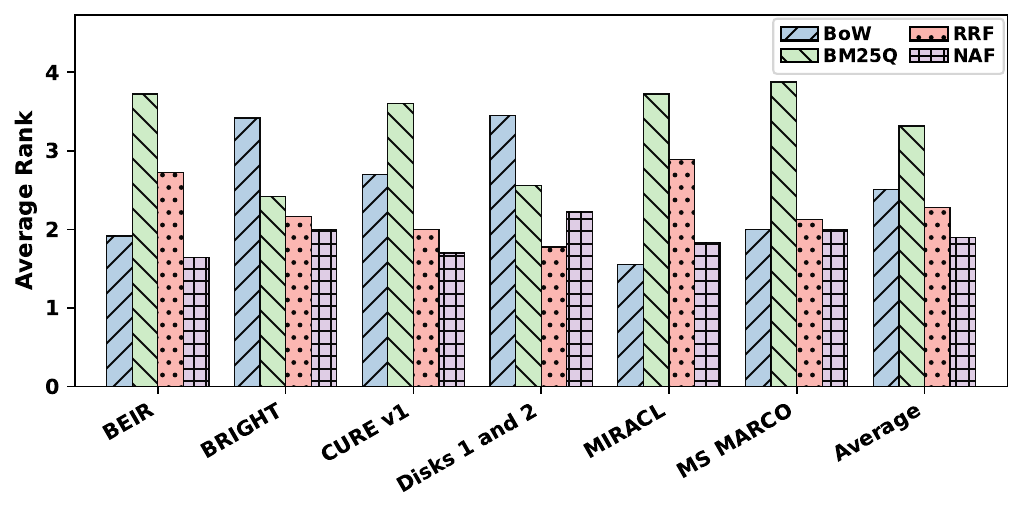}
        \vspace{-0.4cm}
        \caption{nDCG@10}
    \end{subfigure}%
    \hfill
    \begin{subfigure}[t]{\columnwidth}
        \centering
        \includegraphics[width=\columnwidth]{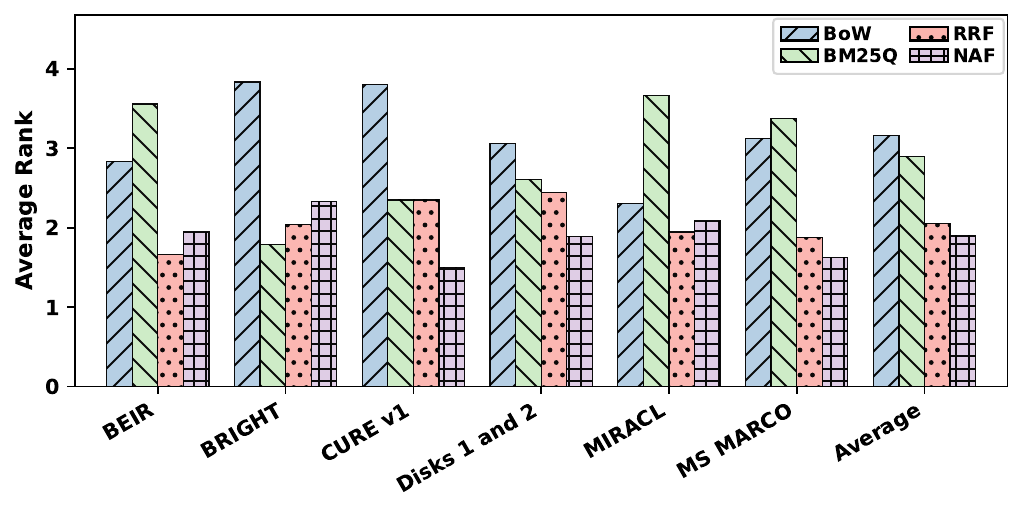}
        \vspace{-0.4cm}
        \caption{Recall@100}
    \end{subfigure}
\vspace{-2mm}
    \caption{Average ranking of BoW, BM25Q, and their fusions (RRF and NAF) across six benchmarks, measured by (a) nDCG@10 and (b) Recall@100.}
\vspace{-2mm}
    \label{fig:average_ranking}
\end{figure}
\section{BM25Q Generalizability}
\autoref{tab:generalizability} reports the average nDCG@10 and Recall@100 for all tasks across six benchmarks: BEIR, BRIGHT, CURE v1, TREC Disks 1 \& 2, MIRACL, and TREC DL19-D23 tracks and their dev sets. BoW and BM25Q yield 35.1 vs.\ 34.3 average nDCG@10, while the fusions are slightly higher (RRF 35.3, NAF 35.5). 
For Recall@100, BM25Q is modestly above BoW (51.0 vs.\ 50.6), and both fusions attain 51.7. Per-suite patterns align with these averages: BM25Q lags behind BoW on nDCG in BEIR, CURE~v1, MIRACL, and TREC DL tracks, but is ahead in the remaining two. For recall, the trend is reversed with BM25Q being ahead in all datasets but BEIR and MIRACL; fusion typically matches or exceeds the better of the two single runs.
Next, we look at the statistical significance of these results using pairwise permutation tests comparing BM25Q, RRF, and NAF to BoW.

As shown in~\autoref{tab:generalizability}, relative to BoW, BM25Q shows a small but statistically significant drop in nDCG@10 across all tasks ($\Delta=-0.8$, 95\% CI $[-1.4,\,-0.2]$, $p=0.005$) and an \textit{NS} change in Recall@100 with $p=0.408$. 
In contrast, fusing BoW and BM25Q is a low-effort win. 
Both RRF and NAF yield a clear recall gain with no nDCG@10 harm overall (Recall@100 $\Delta=+1.1$, $p<0.001$; nDCG@10 \textit{NS}).

\Cref{fig:average_ranking} visualizes these findings by showing, for each retrieval method, its average rank within each benchmark. For each task, the four retrieval methods are ranked from 1 (best) to 4 (worst) based on effectiveness. When methods tie, they are assigned the mean of the tied ranks (e.g., methods tied for ranks 2 and 3 are both assigned rank 2.5). These ranks are then averaged across all tasks in the benchmark.
The figure shows that BoW and BM25Q alternate in rank depending on the evaluation metric, indicating comparable performance. In contrast, both fusion methods achieve consistently lower (i.e., better) average ranks across benchmarks, with NAF producing the strongest overall results.

\section{Conclusion}

This work establishes strong, reproducible baselines for reasoning-intensive retrieval on the BRIGHT benchmark by integrating lexical, dense, sparse, and reasoning-focused retrievers, along with listwise LLM rerankers, into widely used toolkits including Anserini, Pyserini, and RankLLM. These integrations lower the barrier to entry for researchers and provide a transparent foundation for evaluating retrieval pipelines in the emerging retrieval-augmented generation (RAG) setting.

Our analysis reveals several key findings. First, query-side BM25 (BM25Q), although under-documented in prior work, substantially improves lexical retrieval effectiveness on BRIGHT's long queries and is necessary to faithfully reproduce previously reported baselines. However, its benefits do not consistently generalize across standard IR benchmarks. In contrast, simple fusion of BoW and BM25Q---particularly normalized average fusion (NAF)---provides the most robust and generalizable configuration, yielding consistent recall improvements without harming early precision. These results suggest that modest modifications to classical lexical retrieval can remain highly effective even in reasoning-oriented settings.

Second, reasoning-focused retrievers significantly outperform generic dense and sparse retrievers, demonstrating the importance of task-aligned representation learning. Listwise LLM reranking further improves effectiveness, particularly for weaker first-stage retrievers, though its gains diminish as first-stage retrieval quality improves and may even degrade performance when reranker capability is insufficient relative to retriever strength. This highlights the importance of considering the interaction between retrieval and reranking components rather than optimizing them independently.

Finally, our audit of the BRIGHT corpus identifies data quality issues---including duplicate documents, empty passages, and missing relevance labels---that can affect evaluation outcomes. Addressing these issues improves the reliability of effectiveness measurements and underscores the critical role of dataset construction and preprocessing in reasoning-oriented retrieval benchmarks.

Taken together, our contributions provide both practical guidance and reproducible infrastructure for evaluating retrieval systems on reasoning-intensive tasks, and clarify which established retrieval practices remain effective and which require reconsideration in the era of LLM-driven retrieval.

\section{Limitations and Future Work}


\paragraph{Reasoning-Aware Reranking.}
While general-purpose LLMs as listwise rerankers provide substantial gains, specialized reasoning-aware pointwise rerankers trained with supervised fine-tuning or reinforcement learning achieve higher effectiveness on BRIGHT. Integrating such models into RankLLM and systematically studying their training strategies, scaling behavior, and robustness remains an important direction for closing the gap between reproducible references and leaderboard-level performance.

\paragraph{Query-Side and Agentic Retrieval Interventions.}
LLM-based query expansion, rewriting, and iterative retrieval have demonstrated strong potential for reasoning-intensive tasks. Future work should explore adaptive and multi-stage retrieval pipelines that dynamically refine queries based on intermediate results, while carefully evaluating the associated efficiency and latency trade-offs.

\paragraph{Hybrid and Adaptive Lexical–Neural Retrieval.}
Our findings show that simple fusion of lexical variants can provide robust gains, but the optimal combination likely depends on query characteristics, domain, and retriever strength. Future work should investigate adaptive fusion strategies, hybrid dense–sparse–lexical systems, and query-dependent weighting schemes that better leverage complementary retrieval signals.

\paragraph{Dataset Quality and Benchmark Reliability.}
Our corpus audit shows that preprocessing artifacts and incomplete relevance annotations can measurably affect evaluation. Future efforts should focus on improving dataset construction, including duplicate removal, passage segmentation quality, and more comprehensive relevance labeling, to ensure fair and reliable comparisons across systems.


Overall, advancing reasoning-oriented retrieval will require joint progress in retriever design, reranking methods, retrieval pipelines, and benchmark quality. We hope the reproducible references and analyses provided in this work serve as a foundation for continued research in this rapidly evolving area.

\begin{acks}
This research was supported in part by the Natural Sciences and Engineering Research Council (NSERC) of Canada. 
Additional funding was provided by Snowflake and the Institute of Information \& Communications Technology Planning \& Evaluation (IITP) grant funded by the Korean Government (MSIT) (No.\ RS-2024-00457882, National AI Research Lab Project).
\end{acks}

\bibliographystyle{ACM-Reference-Format}
\bibliography{sample-base}

\clearpage



\end{document}